\documentclass[12pt]{iopart}
\usepackage{iopams} 
\usepackage[normalem]{ulem}
\begin{document}
\title[Is there vacuum when there is mass? Solutions for massive gravity]{Is there vacuum when there is mass?\\
Vacuum and non-vacuum solutions for massive gravity}
\author{Prado Mart\'in--Moruno {\rm and} Matt Visser}
\address{School of Mathematics, Statistics, and Operations Research\\
Victoria University of Wellington, PO Box 600\\
Wellington 6140, New Zealand}
\ead{prado@msor.vuw.ac.nz, matt.visser@msor.vuw.ac.nz}
\begin{abstract}
Massive gravity is a theory which has a tremendous amount of freedom to describe different cosmologies;  but at the same time the various solutions one encounters must fulfill some rather nontrivial constraints.  
Most of the freedom comes not from the Lagrangian, which contains only a small number of free parameters (typically 3 depending on counting conventions), but from the fact that one is in principle free to choose the reference metric almost arbitrarily --- which effectively introduces a non-denumerable infinity of free parameters. 
In the current paper we stress that although changing the reference metric would lead to a different cosmological model, this does not mean  that the dynamics of the universe can be entirely divorced from its matter content.
That is, while the choice of reference metric certainly influences the evolution of the physically observable foreground metric, the effect of matter \emph{cannot} be neglected.
Indeed the interplay between matter and geometry can be significantly changed in some specific models; effectively since the graviton would be able to curve the spacetime by itself, without the need of matter.
Thus, even the set of vacuum solutions for massive gravity can have significant structure. 
In some cases the effect of the reference metric could be so strong that no 
conceivable material content would be able to drastically affect the cosmological evolution.

\bigskip
\noindent
28 June 2013; \LaTeX-ed \today.


\bigskip
\noindent
\emph{Dedicated to the memory of Professor Pedro F. Gonz\'alez--D\'iaz}

\end{abstract}

\vspace{2pc}

\clearpage
\section{Introduction}
The theory of massive gravity recently introduced by de Rham, Gabadadze, and Tolley~\cite{deRham:2010kj} 
(see also de Rham and Gabadadze~\cite{deRham:2010ik}), has led to an explosive increase of interest in the possibility of having a graviton with a non-vanishing mass, revitalizing the old topic of $f$--$g$ gravity. 
The principal reason for this renewed interest has been that this theory is free of the otherwise problematic sixth degree of freedom, 
the Boulware--Deser ghost~\cite{Boulware:1973my};  even at the nonlinear level and with a generic reference 
metric~\cite{Hassan:2011tf, Hassan:2011ea} (see also~\cite{Comelli:2012vz}).
Cosmologies with a Minkowski reference metric, which were used to construct the first version of this theory~\cite{deRham:2010kj}, 
have been considered in reference~\cite{D'Amico:2011jj}, where the authors showed that one cannot have a FLRW physical metric compatible 
with such a reference spacetime.  
This result is a direct consequence of the Bianchi-inspired constraint. 
(There is one minor exception using the Milne spacetime as reference metric~\cite{Gumrukcuoglu:2011ew}; 
this is a particular open FLRW universe which can be viewed as a particular chart on a portion of Minkowski spacetime.
The results of references~\cite{D'Amico:2011jj, Gumrukcuoglu:2011ew} about massive gravity with a flat reference metric have been
recovered in references~\cite{Volkov:2011an, Comelli:2011zm}.)
The incompatibility of a physical cosmological space with a Minkowski reference metric had already been suggested in earlier work on massive gravity~\cite{Visser:1997hd}, where the author also pointed out that in massive gravity any arbitrary cosmology can be accommodated in the physical sector by suitably fitting the required reference metric. 
This point has been recently stressed in references~\cite{vonStrauss:2011mq} and~\cite{Baccetti:2012ge}. 

On the other hand, some authors have suggested that one should focus attention only on the particular branch of solutions where the two 
spatial metrics, foreground and background,  are strictly proportional to each other, with a fixed constant proportionality factor~\cite{Gumrukcuoglu:2011zh}. 
This restriction would imply elimination of all nontrivial interaction terms, those which are not equivalent to a cosmological constant, and such a constraint could only be understood if one is trying to preserve continuity in the  parameter space of the theory~\cite{Baccetti:2012bk}.
Moreover, it has been argued that the solutions of this restricted branch might be unstable~\cite{DeFelice:2012mx}. 
It must be pointed out that the original reason of paying attention only to this particular restricted set of solutions, which are equivalent to simply introducing
a cosmological constant,  could be misleading. 

The other branch has been suggested to be trivial~\cite{Gumrukcuoglu:2011zh},  by arguing that the physical metric would be completely determined from the reference metric, through the Bianchi-inspired constraint, and not by the matter content.
Such a conclusion can be extracted only by assuming some simple relation between the cosmic times of the two metrics, since a nontrivial function could radically change the behavior of the physical cosmology.  
This has been pointed out in reference~\cite{Capozziello:2012re} in the case of bimetric gravity (reference~\cite{Nojiri:2012re}
contains a similar study in the framework of $f(R)$ extensions of bimetric gravity~\cite{Nojiri:2012zu}), 
and for massive cosmologies we will explicitly demonstrate the corresponding result in section~\ref{s2}. 
In fact, interesting cosmological solutions with a de Sitter reference metric have recently been presented in 
reference~\cite{Langlois:2012hk}.  
Overall, we shall see that while the choice of reference metric certainly influences  the evolution of the physically observable  foreground metric, the effect of matter \emph{cannot} be neglected.

Following the lead of reference~\cite{Fasiello:2012rw}, where Fasiello and Tolley have studied the classical stability of homogeneous and isotropic solutions of this nontrivial branch, one should of course take into account the possible tension 
between the Vainshtein mechanism~\cite{Vainshtein:1972sx}, (originally introduced to avoid the van~Dam--Veltman--Zakharov (vDVZ) 
discontinuity~\cite{vanDam:1970vg, Zakharov:1970cc}), and the Higuchi bound~\cite{Higuchi} ($\tilde{m}^2 > 2 H^2$). Nonetheless, these solutions should certainly not be excluded from the very beginning.
In order to avoid that tension we will also consider the possibility of having an anti de~Sitter (adS) reference metric~\cite{Kogan:2000uy, Porrati:2000cp}. 
Such a massive gravity model would share some interesting features with that based on assuming a de~Sitter (dS) reference metric;  that is it implies trivial curvature when there is no material content. 
As we show in section~\ref{s3}, this is not a common characteristic of all massive gravity models.
Nevertheless, an adS reference metric cannot successfully describe our own physical universe, at least assuming
that it is homogeneous and isotropic at large scales, see section~\ref{s4}.
The already mentioned conclusions are summarized in section~\ref{s5}, where additional comments are also added.

\section{Matter matters}\label{s2}
In this section we will show that given a particular model of massive gravity, that is, once the reference metric is fixed,
adding matter necessarily changes the dynamics of the universe in the nontrivial cosmological branch. 
To see this, let us consider a homogeneous and isotropic reference metric
\begin{equation}\label{metricf}
ds_f^2 = - d\tau^2 + b(\tau)^2 \left[ {dr^2\over 1- k r^2} + r^2 \left(d\theta^2 + \sin^2\theta\; d\phi^2\right) \right],
\end{equation}
to agree with the symmetry of our physical cosmological spacetime
\begin{equation}\label{metricg}
ds_g^2 = - dt^2 + a(t)^2 \left[ {dr^2\over 1- k r^2} + r^2 \left(d\theta^2 + \sin^2\theta\; d\phi^2\right) \right],
\end{equation}
where we have assumed that both spacetimes have the same sign of spatial curvature.
We allow an arbitrary relation $\tau=\tau(t)$ between the two time coordinates, 
thereby excluding particular solutions 
which would not be compatible with both metrics being diagonal in the same coordinate patch. 
Thus, we can express the relation between the two cosmic times as~\cite{Capozziello:2012re}
\begin{equation}\label{tau}
d\tau=\pm N(t) \; dt.
\end{equation}
Here $N(t)>0$;  we have explicitly written the two possible signs instead of absorbing them into $N(t)$.

It is already well known~\cite{vonStrauss:2011mq, Baccetti:2012ge, Capozziello:2012re}, that the modified Friedmann equation 
describing the dynamics of the universe in this model can be written as
\begin{equation}\label{Fried}
H^2+\frac{k}{a^2}=\frac{m^2}{3}\rho_\mathrm{g}+\frac{8\pi\,G}{3}\rho_\mathrm{m}+\frac{\Lambda}{3},
\end{equation}
where the effects due to a non-vanishing graviton mass have been described as an effective fluid with $\rho_\mathrm{g}$ given 
by~\cite{Baccetti:2012ge}
\begin{equation}\label{rho}
 \rho_\mathrm{g}=\frac{b}{a}\left(3\beta_1+3\beta_2\; \frac{b}{a}+\beta_3\; \frac{b^2}{a^2}\right).
\end{equation}
Here $\beta_1$, $\beta_2$, and $\beta_3$ are parameters of the theory; while $\rho_\mathrm{m}$ is the standard matter content 
filling the universe, which obeys the usual conservation equation,
\begin{equation}\label{materia}
\dot\rho_\mathrm{m}+3H\,[1+w_\mathrm{m}(a)]\,\rho_\mathrm{m}=0; \qquad w_\mathrm{m}(a) = p_\mathrm{m}(a)/\rho_\mathrm{m}(a);
\end{equation}
and $\Lambda$ includes the common cosmological constant term plus some terms coming from the theory of massive gravity~\cite{Baccetti:2012bk}.
The effective fluid which appears in the modified Friedmann equation is also conserved~\cite{Baccetti:2012ge}, and the corresponding conservation equation leads to the nontrivial part of the Bianchi-inspired constraint.  
Expressing \; $\dot{} \equiv d/dt$ and $'\equiv d/d\tau$, it can be written as~\cite{Capozziello:2012re}
\begin{equation}\label{consev}
\dot b(t)=N(t) \, \dot a(t), \qquad\hbox{which is equivalent to} \qquad  \dot a(t(\tau))=b'(\tau).
\end{equation}
Taking this expression into account in the modified Friedmann equation (\ref{Fried}), and then simplifying, we have
\begin{equation}\label{pol}
\fl
\frac{8\pi\,G}{3}\rho_\mathrm{m}(a)\, a^2+\frac{\Lambda}{3}a^2+3\beta_1\, b(\tau)\, a+3\beta_2\, b^2(\tau)+\beta_3\, \frac{b^3(\tau)}{a}-b'^2(\tau)-k=0.
\end{equation}
Here we have explicitly shown the dependence on $\rho_\mathrm{m}=\rho_\mathrm{m}(a)$ due to equation~(\ref{materia}).
Equation~(\ref{pol}) can be seen as an implicit algebraic (non-derivative) equation for $a$, with the quantity $\rho_\mathrm{m}(a)$ furthermore implicitly depending on $w_\mathrm{m}(a)$. 
(In the specific case that $w_\mathrm{m}(a)= w_\mathrm{m}$ is a constant, which is often a good approximation for significant epochs of cosmological time, equation~(\ref{pol}) can more specifically be seen as a polynomial equation for $a$, with $\rho_\mathrm{m}\propto a^{-3(1+w_\mathrm{m})}$, and so with the degree of the polynomial depending on $w_\mathrm{m}$.)
In either case, the particular function $a(\tau)$ can only be obtained by specifying $w_\mathrm{m}(a)$, so we shall write it as
$a_\mathrm{w}(\tau)$ to make this point explicit. 
Explicitly, there will be some function $A(\_\_,\_\_)$ such that
\begin{equation}
a_\mathrm{w}(\tau) = A\left( b(\tau), b'(\tau) \right).
\end{equation}
Differentiating $a_\mathrm{w}(\tau)$ with respect to $\tau$, we can now obtain the function $t_\mathrm{w}(\tau)$ through
\begin{equation}\label{t}
t_\mathrm{w}(\tau)=\int d\tau\;\frac{a_\mathrm{w}'(\tau)}{b'(\tau)}, 
\end{equation}
where we have used equation~(\ref{consev}) to get expression~(\ref{t}). Note the implicit dependence on $w_\mathrm{m}(a)$ and hence on 
the matter content. 

Thus, despite the fact that the function $a(\tau)$ is determined (up to an integration constant) by the reference metric through the Bianchi-inspired constraint (\ref{consev}), the influence of the matter content cannot be neglected --- matter matters. 
Furthermore, the behavior of the observable foreground universe is described by $a(t)$, which can describe an evolution completely different to that given by $a(\tau)$, and which can be obtained only once the matter content is known.

On the other hand, one could proceed the other way around and reconstruct any particular $a(t)$ by carefully fixing $b(\tau)$ and $w_\mathrm{m}(a)$, thus, any cosmology could be described by changing the reference metric. Therefore, it would be desirable to have some additional criteria to choose a particular reference metric, so that the theory maintains its predictive character.

\section{Gravity without matter}\label{s3}

If one now considers a massive gravity theory suitable for describing our own physical universe, it is natural to assume that the reference 
metric used to define the theory would be homogeneous and isotropic. 
This particular theory, with that fixed reference metric, should also be consistent in other situations, for example in vacuum. 
So, in order to better understand the implications of the various models of massive gravity, let us go back to 
the modified Friedmann equation~(\ref{Fried}), which in the absence of matter can be simplified as
\begin{equation}\label{Friedvacio}
\frac{b'^2+k}{b^2}=\frac{\Lambda}{3}\left(\frac{a}{b}\right)^2+3\beta_1\frac{a}{b}+3\beta_2+\beta_3\left(\frac{a}{b}\right)^{-1}.
\end{equation}
This equation is a cubic polynomial in $a$, and a first-order differential equation in $b(\tau)$, which has different implications depending on the symmetry of the reference metric.

\subsection{Constant curvature reference metrics (dS/Minkowski/adS)}
As is well known (see for example, reference~\cite{HyE}) both the dS and adS spacetimes can be written using FLRW charts;  open, flat, or closed for the former space, but necessarily open for the latter. 
In either of these cases one would have
\begin{equation}\label{eq}
\frac{b'^2+k}{b^2}=\lambda,
\end{equation}
with $\lambda>0$ for dS and $\lambda<0$ for adS. (Minkowski space corresponds to $\lambda=0$ and $k=0$, 
and has already been discussed in the introduction.) 
It must be noted that we are not imposing any dynamic Friedmann 
equation for the reference metric; equation~(\ref{eq}) is a purely algebraic equality fulfilled for the particular $b(\tau)$ of these background spaces.

Thus, in both these cases equation~(\ref{Friedvacio}) is simply a constant coefficient polynomial in $x$, with $x\equiv a/b$. 
So, the solution is
\begin{equation}
a(\tau)=C(\Lambda,\beta_1,\beta_2,\beta_3,\lambda)\,b(\tau),
\end{equation}
with the constant $C(\Lambda,\beta_1,\beta_2,\beta_3,\lambda)$ corresponding to one of the positive roots of the 
the cubic equation
\begin{equation}\label{cubic}
\frac{\Lambda}{3}x^3+3\beta_1\,x^2+3\left(\beta_2-\lambda\right)\,x+\beta_3=0,
\end{equation}
which we shall not write out explicitly to keep the analysis as clean as possible.
In the case that there is no real positive solution of the cubic, then there is no well-defined physical metric for those particular values of the parameters
$\Lambda$, $\beta_1$, $\beta_2$, $\beta_3$, and $\lambda$.
Since any such solution implies $a'(\tau)=C\,b'(\tau)$, then equation~(\ref{t}) leads to $t=C\tau+D$, where without loss of generality we can set $D\to0$. 
Therefore, 
\begin{equation}\label{sol-adS}
 a(t)=C\,b(\tau(t))=C\,b\left(t/C\right).
\end{equation}
Thus, the solution of the physical metric in vacuum is a spacetime of the same type as the reference metric, 
that is dS or adS, with an effective cosmological constant $\Lambda_\mathrm{eff}=\Lambda_0/C^2$, with $\Lambda_0$ the cosmological constant corresponding to the reference metric  --- thus it is also a vacuum solution.
The high symmetry of these particular reference spaces is so constraining that the only solution (in vacuum) for the physical space must be compatible
with that same symmetry, with the only effect of the theory being the change of  the value of the cosmological constant with respect to that of the reference spacetime. 
The introduction of matter, of course, would change this situation, since $a(\tau)=C\,b(\tau)$ would no longer be a general solution for constant $C$.

\subsection{FLRW reference metrics}
In the case of a generic FLRW reference metric, equation~(\ref{Friedvacio})
would not simplify so beautifully, since we will now have
\begin{equation}
\frac{b'^2+k}{b^2}=f(\tau).
\end{equation}
In this case the cubic equation implied by equation~(\ref{Friedvacio}) has a coefficient which is now a function of time. 
To be more explicit about this, note that the relevant cubic is now
\begin{equation}\label{cubic2}
\frac{\Lambda}{3}x^3+3\beta_1\,x^2+3\left(\beta_2-f(\tau)\right)\,x+\beta_3=0.
\end{equation}
Let the positive real roots (if any) be denoted $C(\tau) = C(\Lambda,\beta_1,\beta_2,\beta_3,f(\tau))$ then 
\begin{equation}
a(\tau)=C(\Lambda,\beta_1,\beta_2,\beta_3,f(\tau))\,b(\tau),
\end{equation}
The effective fluid which we have defined through the energy density~(\ref{rho}) would not be equivalent to a cosmological
constant, since now $\rho(t) =\rho(a(\tau))=\rho(\tau)$ is time dependent. That will, lead at the end of the day, to a modified Friedmann equation:
\begin{equation}
H^2+\frac{k}{a^2}=\frac{m^2}{3}\rho(t)+\frac{\Lambda}{3}.
\end{equation}
Therefore, in the nontrivial branch of solutions a non-vanishing graviton mass would generally curve the foreground spacetime 
in some nontrivial manner even in vacuum, leading to a situation where there are gravitational effects evolving in time in the absence of matter.
In some sense, there would be gravitation even without matter.
The old formal question, dating back to the ancient Greeks,  concerning the possible existence of space itself in the absence
of matter would here be changed to the possible existence of nontrivial gravitational effects without material content.
The dS and adS spaces would be the only consistent reference metrics from that point of view, since the massive graviton in those models does not curve the spacetime in the absence of matter in any nontrivial way, (merely adjusting the value of the constant curvature). 
That is, in both the dS and adS massive gravity models the effects of the non-vanishing graviton mass would become evident only in the presence of matter, hiding as a simple cosmological constant contribution in any vacuum situation.

\section{Reference adS protects against acceleration}\label{s4}

Following the spirit of the previous section, we shall now consider if an adS spacetime can be the reference metric leading to a model of massive gravity capable of describing our universe. 
(Models with a dS reference metric have been considered in references~\cite{Langlois:2012hk,Fasiello:2012rw}.)
We have just seen that the vacuum solution for the physical metric in a model with an adS reference metric is an adS spacetime, equation~(\ref{sol-adS}). 
Thus, this solution would have both a phase of decelerated expansion and a contracting phase. 
The addition of a matter component to equation~(\ref{Friedvacio}) would then decelerate the expansion phase, (and accelerate the contracting one), at least if this matter fulfills the usual energy conditions~\cite{twilight}. 
(In particular, if the matter component satisfies the strong energy condition, SEC~\cite{science, cosmo99, minimal}.)
Thus, the model would not be suitable for describing our universe, at least if the matter content fulfills the energy conditions.

One might think that the situation could be improved by increasing the value of $\Lambda$ in equation~(\ref{Friedvacio}), but this is not  the case since one would still have a linear relation between the scale factors, as it is shown by equation~(\ref{cubic}).
In order to show this in a simple way, we fix some of the parameters of the model.  
For example, set $\beta_1=-1$, $\beta_2=\beta_3=0$, and take into account that $k=-1$ in order to be compatible with the adS reference metric. 
Then we have $a=C\,b$ and $t=C\tau+D$, with
\begin{equation}
 C=3\;\frac{3+\sqrt{9-4\Lambda/3}}{2\Lambda}.
\end{equation}
By increasing the value of $\Lambda$ we eventually obtain a complex value of $C$, so the solution is not well defined.
The adS reference space protects the physical space against acceleration in two ways; not only by inducing a deceleration, but also an adS reference metric is not
compatible with well defined solutions for large values of a positive cosmological constant. This result is a consequence of the linearity between the scale factors in vacuum and, as we have argued above, it cannot be changed by adding matter which fulfils the SEC. 

Therefore, we can conclude that there is no accelerating homogeneous and isotropic solutions for models with an adS reference metric in the absence of matter violating the energy conditions, at least if we assume that the metrics take the form given by equations~(\ref{metricf}) and (\ref{metricg}), which means that both metrics have the same spatial curvature and can be written in a diagonal way in the same coordinate patch.

\section{Discussion}\label{s5}

As we have emphasized in this paper: In massive cosmologies there is no reason to restrict our attention only to that branch of solutions
where both scale factors are strictly proportional with a fixed constant of proportionality leading to a situation which is equivalent to a cosmological constant. 
Although there is a great deal of freedom in the set of solutions to this theory, once we have restricted to a particular model with a given reference metric, the dynamics of the physical universe would of course be determined by the matter content of that universe. 
The reference metric will certainly affect the particular solutions, but so will do the matter content.

On the other hand,  we have shown that in those models of massive gravity with dS and adS reference metrics all the vacuum solutions behave as the solutions of the trivial branch, implying strictly proportional scale factors and cosmic times. 
This result can be understood by observing that the graviton should hide the effect of a non-vanishing mass in vacuum, in order to avoid the presence of nontrivial gravitational effects in the absence of matter.
If one were tempted to follow that line of thinking, he/she would conclude that these are the only consistent massive gravity models.
As the models with dS reference metric (as well as those with FLRW reference metric) have been suggested to be affected by some tension between the Higuchi bound and the Vainshtein mechanism for this branch of solutions even when the background metric and the reference metric are distinct (which leads to metric fluctuations which are not of the Fierz-Pauli form) \cite{Fasiello:2012rw}, a promising possibility is to consider an adS reference metric, where those particular problems are not present~\cite{Kogan:2000uy, Porrati:2000cp}.
Nevertheless, as we have demonstrated in this paper, the attractive effect induced by such a reference metric would be so strong that it cannot
describe the current accelerated expansion of our universe, even if we were to introduce a positive cosmological constant with large value.

The non-existence of homogenous and isotropic accelerating solutions for models with an adS reference metric is a general result which is valid for any model where both metrics can be written in a diagonal way in the same coordinate patch, having the same spatial curvature ($k=-1$), and assuming the absence of matter violating the energy conditions.
The dynamics of the observable physical metric is determined by both the reference metric and the material content, and, therefore, the visible causal structure is affected by the reference metric. In the case of an adS reference metric this effect is strong enough to forbid the existence of accelerating solutions (for any matter content). 
For other cosmological models fulfilling the diagonal assumption, the causal structure of the physical metric can be studied by noting that this assumption is a particular case of the generalized Gordon ansatz (as introduced and discussed in Ref.~\cite{Baccetti:2012ge}). Thus, in view of the FLRW symmetries common to both metrics the light cones of both metrics have the same axes, those of the physical metric being inside or outside the light cones of the reference metric depending on the particular solution obtained when the material content is considered. 

It should be stressed that we are not suggesting that there are not massive cosmologies with FLRW reference metrics able to describe our Universe. In fact, we have proven that any universe can be described by massive gravity if one were to carefully choose an appropriate reference metric. Nevertheless, on one hand, the resulting universe would be affected by the already mentioned Higuchi-Vainstein tension \cite{Fasiello:2012rw}, giving not a lot of chance for a stable universe in which general relativity can be recovered at some range, and, on the other hand, one should accept the presence of nontrivial gravitational effects in vacuum.

Finally, it should be pointed out that a similar result concerning nontrivial curvature in the absence of material content for some reference metrics \emph{cannot} be
extracted in the framework of ghost-free bimetric gravity~\cite{Hassan:2011zd}.
In bimetric gravity, if we obtain an effective fluid which is not equivalent to a cosmological constant in some gravitation sector where there is no material content, then all we can say for sure that there would be some material content in the other gravitational sector. 
That is, some material content is automatically present in the theory curving the spacetime, even if that material content cannot be directly observed in
the foreground sector.
This reinforces the idea that, in spite of the mathematical similarities, massive gravity and bimetric gravity should be considered as conceptually very different theories~\cite{Baccetti:2012bk}.
It must be emphasised that we are indicating that an effective fluid inequivalent to a cosmological constant is possible in bimetric gravity only if there is some material content, but not the negation of this statement. 
We can of course quite easily have an effective fluid equivalent to a cosmological constant in the presence of matter, which would (as has been long known~\cite{Gurses:1980as,Gurses:1981an} and has been recently  pointed out in the ghost-free theory~\cite{Baccetti:2012re}) only require that the effective fluid of the other gravitational sector must also be equivalent to a cosmological constant.

\ack
PMM wishes to dedicate this article to the memory of Pedro F. Gonz\'alez--D\'iaz, whose views about physics were and will ever be an inspiration.
\\
PMM acknowledges financial support from a FECYT postdoctoral mobility contract of the Spanish Ministry of Education through National Programme No. 2008-2011.
\\
MV acknowledges support via the Marsden Fund and via a James Cook Fellowship, both administered by the Royal Society of New Zealand. 

\vspace{12pt}
\section*{References}


\end{document}